\documentclass[twocolumn,showpacs,preprintnumbers,amsmath,amssymb, superscriptaddress, unsortedaddress,floatfix]{revtex4}

\usepackage{amsmath}
\usepackage{graphicx}
\usepackage{bm}

\newcommand{\ket}[1]{\mbox{$|#1\rangle$}}

\begin{document}

\title{Methods for linear optical quantum Fredkin gate}

\author{Yan-Xiao Gong}
\email{yxgong@mail.ustc.edu.cn} \affiliation{Centre for Quantum
Computer Technology, Department of Physics, University of
Queensland, St Lucia 4072, Australia} \affiliation{Key Laboratory of
Quantum Information, University of Science and Technology of China,
CAS, Hefei, 230026, People's Republic of China}
\author{Guang-Can Guo}
\affiliation{Key Laboratory of Quantum Information, University of
Science and Technology of China, CAS, Hefei, 230026, People's
Republic of China}
\author{Timothy C. Ralph}
\affiliation{Centre for Quantum Computer Technology, Department of
Physics, University of Queensland, St Lucia 4072, Australia}

\date{\today }

\begin{abstract}

We consider the realization of quantum Fredkin gate with only linear
optics and single photons. First we construct a heralded Fredkin
gate using four heralded controlled-not (CNOT) gates. Then we
simplify this method to a post-selected one utilizing only two CNOT
gates. We also give a possible realization of this method which is
feasible with current experimental technology. Another post-selected
scheme requires time entanglement of the input photons but needs no
ancillary photons.
\end{abstract}


\maketitle

\section{Introduction}

Quantum computing \cite{nielsen2000}, due to its potential to solve
problems far beyond classical computers, has attracted great
attention in the last years. Many physical systems have been
considered for a quantum computer \cite{spiller2005}. One promising
system is to use single photons, showing benefits such as low
decoherence and easy single-qubit manipulation. However, such system
suffers a major disadvantage --- the lack of interaction between
individual photon qubits, which is needed for implementing
non-trivial multi-qubit gates. Surprisingly, Knill, Laflamme, and
Milburn have demonstrated that scalable quantum computing was
possible using linear optical elements, single photons and photon
detection \cite{KLM2001}. After that, lots of researches have been
devoted to linear optical quantum computing (LOQC) \cite{Kok2007}.

Here we focus on the implementation of linear optical Fredkin gate
\cite{fredkin1982}, which plays an important role in both classical
computing and quantum computing \cite{nielsen2000}. The Fredkin gate
is also known as a three-qubit controlled-swap gate, that is, if the
control qubit is in state $\ket{1}$, the two target qubits swap
their states and otherwise, they remain in their initial states if
the control qubit is in state $\ket{0}$. In the context of universal
quantum computer, multi-qubit gates are usually thought to be built
by a combination of single- and two-qubit gates. Smolin and
DiVincenzo have shown that five two-qubit gates are sufficient to
implement the Fredkin gate \cite{Smolin1996}. Assuming that the
two-qubit gate, as the controlled-not (CNOT) gate in
\cite{PittmanCNOT}, can be implemented using two ancillary photons
with success probability of $1/4$, their gate needs ten ancillary
photons and the total success probability is
\mbox{$4^{-5}\approx1.0\times10^{-3}$}. It is too difficult and not
possible with current experimental technology. Recently, another
scheme was proposed in \cite{fiurasek_fredkin} by simulating the
Kerr medium in Milburn's optical Fredkin gate \cite{Milburn_Fredkin}
with linear optical elements. It needs only six ancillary photons
with the success probability $4.1\times10^{-3}$.

Recently, the complexity of the Toffoli gate was highly reduced and
the success probability was improved \cite{ralph_Toffoli}. In this
paper, we wish to see if a similar effect can be achieved by
applying those techniques to the Fredkin gate. We propose some
methods for implementing the Fredkin gate with linear optics and
single photons. The qubits in our schemes are all encoded in
polarization states of single photons, so that
\mbox{$\ket{0}\equiv\ket{H}$} and \mbox{$\ket{1}\equiv\ket{V}$},
where $\ket{H}$ ($\ket{V}$) denotes the horizontal (vertical)
polarization state. The rest of the paper is organized as follows.
In the next section we propose a heralded Fredkin gate using four
heralded CNOT gates. In Section \ref{Sec_postselected} we give a
post-selected Fredkin gate, i.e., working in the coincidence basis,
and we also present a possible optical realization which is feasible
with existing technology. In Section \ref{Sec_time} we replace the
four heralded CNOT gates in the heralded scheme with four
post-selected CNOT gates assisted by time entanglement but without
ancillary photons. We conclude in Section \ref{Sec_conclusion}.

\section{Heralded Fredkin gate}\label{Sec_heralded}

\begin{figure}
\centering
\includegraphics[width=8cm]{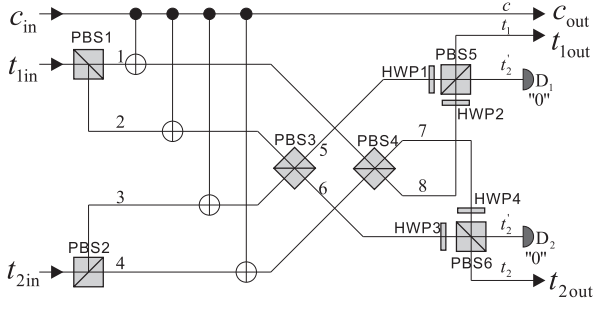}
\caption{Schematic of the heralded linear optical Fredkin gate
comprising four heralded CNOT gates. Lowercase letters and numbers
label the beams. The input control qubit is in beam $c_{\text{in}}$
and the two input target qubits are in beams $t_{1\text{in}}$ and
$t_{2\text{in}}$. Polarizing beam splitters (PBS) transmit
horizontally polarized photons and reflect vertically polarized
photons. Half-wave plates HWP1 and HWP3 are oriented at
$67.5^{\circ}$. HWP2 and HWP4 are set to $22.5^{\circ}$. The gate
succeeds if the two photon number resolving detectors D1 and D2
detect no photons. The output control and target states lie in modes
$c$, $t_1$ and $t_2$.}\label{fig_fourCNOT}
\end{figure}

Our heralded Fredkin gate is built up from four CNOT gates. The
schematic structure is shown in Fig.~\ref{fig_fourCNOT}. To show how
the scheme works, we consider an arbitrary input state written as,
\begin{align}\label{input}
&a_1\ket{H}_{c_{\text{in}}}\ket{H}_{t_{1\text{in}}}\ket{H}_{t_{2\text{in}}}+a_2\ket{H}_{c_{\text{in}}}\ket{H}_{t_{1\text{in}}}\ket{V}_{t_{2\text{in}}}\nonumber\\
+&a_3\ket{H}_{c_{\text{in}}}\ket{V}_{t_{1\text{in}}}\ket{H}_{t_{2\text{in}}}+a_4\ket{H}_{c_{\text{in}}}\ket{V}_{t_{1\text{in}}}\ket{V}_{t_{2\text{in}}}\nonumber\\
+&a_5\ket{V}_{c_{\text{in}}}\ket{H}_{t_{1\text{in}}}\ket{H}_{t_{2\text{in}}}+a_6\ket{V}_{c_{\text{in}}}\ket{H}_{t_{1\text{in}}}\ket{V}_{t_{2\text{in}}}\nonumber\\
+&a_7\ket{V}_{c_{\text{in}}}\ket{V}_{t_{1\text{in}}}\ket{H}_{t_{2\text{in}}}+a_8\ket{V}_{c_{\text{in}}}\ket{V}_{t_{1\text{in}}}\ket{V}_{t_{2\text{in}}},
\end{align}
where $a_i (i=1,2,\dots,8)$ is an arbitrary complex number
satisfying normalization condition.

First the polarizing beam splitter PBS1 (PBS2) transmits the
horizontally polarized photons to beam 1 (4) and vertically
polarized photons to beam 2 (3). Then the photons in each of the
beams 1, 2, 3, and 4 suffer a CNOT gate controlled by the control
state. Therefore, the input state becomes
\begin{align}\label{afterCNOTs}
&a_1\ket{H}_c\ket{H}_1\ket{H}_4+a_2\ket{H}_c\ket{H}_1\ket{V}_3+a_3\ket{H}_c\ket{V}_2\ket{H}_4\nonumber\\+&a_4\ket{H}_c\ket{V}_2\ket{V}_3
+a_5\ket{V}_c\ket{V}_1\ket{V}_4+a_6\ket{V}_c\ket{V}_1\ket{H}_3\nonumber\\+&a_7\ket{V}_c\ket{H}_2\ket{V}_4+a_8\ket{V}_c\ket{H}_2\ket{H}_3.
\end{align}
Next the photons in modes 2 (1) and 3 (4) are mixed at PBS3 (PBS4),
followed by half-wave plates (HWP). Of these, HWP1 and HWP3 oriented
at $67.5^{\circ}$ induce the transformations,
\begin{align}
\label{67.5H}\ket{H}\rightarrow&\frac{1}{\sqrt{2}}\left(-\ket{H}+\ket{V}\right),\\
\label{67.5V}\ket{V}\rightarrow&\frac{1}{\sqrt{2}}\left(\ket{H}+\ket{V}\right),
\end{align}
while HWP2 and HWP4 set to $22.5^{\circ}$ result in,
\begin{align}
\label{22.5H}\ket{H}\rightarrow&\frac{1}{\sqrt{2}}\left(\ket{H}+\ket{V}\right),\\
\label{22.5V}\ket{V}\rightarrow&\frac{1}{\sqrt{2}}\left(\ket{H}-\ket{V}\right).
\end{align}
Finally, PBS5 (PBS6) combines the photons in modes 5 (4) and 8 (7).
Thus, conditioned on a simultaneous zero detection in each of the
modes $t_1'$ and $t_2'$ we can obtain the successful output state in
modes $c$, $t_1$ and $t_2$,
\begin{align}\label{output}
&a_1\ket{H}_c\ket{H}_{t_1}\ket{H}_{t_2}+a_2\ket{H}_c\ket{H}_{t_1}\ket{V}_{t_2}\nonumber\\+&a_3\ket{H}_c\ket{V}_{t_1}\ket{H}_{t_2}+a_4\ket{H}_c\ket{V}_{t_1}\ket{V}_{t_2}\nonumber\\
+&a_5\ket{V}_c\ket{H}_{t_1}\ket{H}_{t_2}+a_6\ket{V}_c\ket{V}_{t_1}\ket{H}_{t_2}\nonumber\\+&a_7\ket{V}_c\ket{H}_{t_1}\ket{V}_{t_2}+a_8\ket{V}_c\ket{V}_{t_1}\ket{V}_{t_2}.
\end{align}

If we use the heralded CNOT gate proposed by Pittman \emph{et al.}
\cite{PittmanCNOT}, we need eight ancillary photons and the success
probability is \mbox{$4^{-5}\approx1.0\times10^{-3}$}. Compared with
the scheme by Smolin and DiVincenzo \cite{Smolin1996}, our scheme
has the same success probability but needs less ancillary photons.
However, our scheme is not as good as Fiur\'{a}\v{s}ek's scheme
\cite{fiurasek_fredkin}. Furthermore, as we shall see, our scheme
can be simplified to a post-selected gate which may be realized with
existing experimental technology.

\section{Post-selected Fredkin gate using two CNOT gates}\label{Sec_postselected}

\begin{figure}
\centering
\includegraphics[width=8cm]{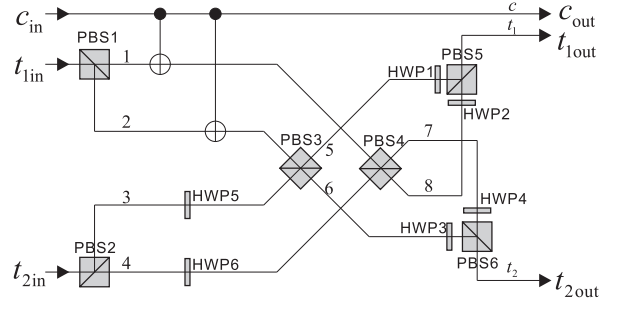}
\caption{Schematic of the post-selected linear optical Fredkin gate
using two controlled-not (CNOT) gates. Lowercase letters and numbers
label the beams. The control and two target qubits are input in
beams $c_{\text{in}}$, $t_{1\text{in}}$ and $t_{2\text{in}}$.
Polarizing beam splitters (PBS) transmit horizontally polarized
photons and reflect vertically polarized photons. Half-wave plates
HWP1, HWP3 and HWP5 are oriented at $67.5^{\circ}$. HWP2, HWP4 and
HWP6 are set to $22.5^{\circ}$. This gate succeeds conditioned on
exact one photon in each of the output beams $c$, $t_1$ and $t_2$.}
\label{fig_twoCNOT}
\end{figure}

We now consider the construction of a post-selected gate. By this we
mean that a gate succeeds conditioned on simultaneous successful
detection of exact one photon for each qubit, so-called coincidence
detection. Fig.~\ref{fig_twoCNOT} is the schematic of a
post-selected Fredkin gate. Comparing this scheme with the heralded
one shown in Fig.~\ref{fig_fourCNOT}, we can see that the
simplification is replacing the two CNOT gates implementing on the
photons in beams 3 and 4 controlled by the photon in beam $c$ by
HWP5 ($67.5^{\circ}$) and HWP6 ($22.5^{\circ}$) with the
transformations given by Eqs.~(\ref{67.5V}) and (\ref{22.5H}),
respectively.

Therefore, for the input state given by Eq.~(\ref{input}), the state
before PBS3 and PBS4 is
\begin{gather}\label{aftertwoCNOTs}
\left(a_1\ket{H}_c\ket{H}_1+a_3\ket{H}_c\ket{V}_2+a_5\ket{V}_c\ket{V}_1+a_7\ket{V}_c\ket{H}_2\right)\nonumber\\
\otimes\frac{1}{\sqrt{2}}\left(\ket{H}_4+\ket{V}_4\right)+\left(a_2\ket{H}_c\ket{H}_1+a_4\ket{H}_c\ket{V}_2\right.\qquad\ \nonumber\\
\left.+a_6\ket{V}_c\ket{V}_1+a_8\ket{V}_c\ket{H}_2\right)\otimes\frac{1}{\sqrt{2}}\left(\ket{H}_3+\ket{V}_3\right).
\end{gather}
Then through the analogy analysis in Section \ref{Sec_heralded} and
in the case of coincidence detection of the output modes $c$, $t_1$
and $t_2$, we can obtain the success output state the same as
Eq.~(\ref{output}).

\begin{figure}
\centering
\includegraphics[width=8.5cm]{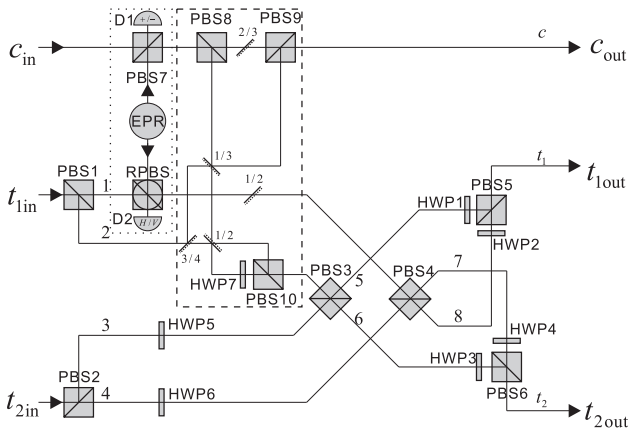}
\caption{Optical realization of a post-selected Fredkin gate.
Lowercase letters and numbers label the beams. Photons in beams
$c_{\text{in}}$, $t_{1\text{in}}$ and $t_{2\text{in}}$ are the
control and two target qubits. Polarizing beam splitters (PBS)
transmit horizontally polarized ($\ket{H}$) photons and reflect
vertically polarized ($\ket{V}$) photons. A heralded controlled-not
(CNOT) gate proposed in \cite{PittmanCNOT} lies in the dotted box.
EPR is an ancillary Bell state source. RPBS, transmitting
$45^{\circ}$ polarized ($\ket{+}$) photons and reflecting
$-45^{\circ}$ polarized ($\ket{V}$) photons, can be realized by
inserting one half-wave plate (HWP) oriented at $22.5^{\circ}$ in
each of the two inputs and two outputs of a PBS. D1 and D2 are
photon detectors detecting $+/-$ basis and $H/V$ basis,
respectively. A post-selected CNOT gate suggested in
\cite{RalphCNOT} is enclosed by the dashed box. Beam splitters are
represented as black lines with their reflectivity indicated aside
and dotted line indicates the surface from which a sign change
occurs upon reflection. HWP1, HWP3 and HWP5 are oriented at
$67.5^{\circ}$. HWP2, HWP4 and HWP6 are set to $22.5^{\circ}$. HWP7
is set to $45^{\circ}$. This gate succeeds conditioned on the
coincidence of the successful detection at D1 and D2 and exactly one
photon in each of the output beams $c$, $t_1$ and $t_2$.}
\label{fig_twoCNOTdetail}
\end{figure}

Fig.~\ref{fig_twoCNOTdetail} shows a possible optical realization of
this scheme. We utilize a heralded CNOT gate proposed in
\cite{PittmanCNOT}, with the success probability $1/4$. Another CNOT
gate need not be heralded and a post-selected CNOT gate given in
\cite{RalphCNOT} can work with the success probability $1/9$.
However, as in our scheme the target state of the second CNOT gate
is known ($\ket{V}$ or vacuum), it turns out that the gate can be
optimized for maximum success probability $1/6$ \cite{ralph_scaling,
ralph_Toffoli} (see the gate in the dashed box). Therefore in the
case of fivefold coincidence, i.e., detection of exact one photon in
each of the output modes $c$, $t_1$ and $t_2$ and successful
detection at D1 and D2, the gate succeeds with a total probability
of success
\mbox{$1/4\times1/6\times1/8=1/192\approx5.2\times10^{-3}$}. As an
ancillary Bell state is needed, to implement this scheme requires at
least a five-photon source, which is available at present
\cite{zhao_five, Zhang_six, lu_graph}, and therefore our scheme is
feasible with current technology. However, the low success
probability of our scheme would make the experiment more difficult
and longer time detection would be needed.

\section{Post-selected Fredkin gate assisted by time entanglement}\label{Sec_time}

\begin{figure}[tb]
\centering
\includegraphics[width=5cm]{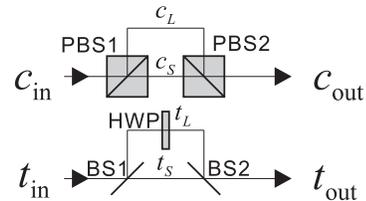}
\caption{Schematic of the controlled-not (CNOT) gate demonstrated in
\cite{Sanaka_CNOT}. Lowercase letters label the beams. Polarizing
beam splitters (PBS) transmit horizontally polarized photons and
reflect vertically polarized photons. The control (target) photon is
input in mode $c_{\text{in}}$ ($t_{\text{in}}$) and output is in
mode $c_{\text{out}}$ ($t_{\text{out}}$). BS1 and BS2 are both
balanced beam splitters. HWP is a half-wave plate set to
$45^{\circ}$.} \label{fig_CNOT}
\end{figure}

In this section we introduce another post-selected Fredkin gate
assisted by time entanglement. Let us first remind the reader of the
CNOT gate presented by Sanaka \emph{et al.} \cite{Sanaka_CNOT} (see
Fig.~\ref{fig_CNOT}).

The control and target photons are a photon pair generated by
spontaneous parametric down-conversion pumped by a Continuous Wave
(CW) laser. Such a source is said to be time-energy entangled
\cite{Franson_time} as the photon pair is in a superposition of many
possible emission times. The control photon is split along the short
($c_L$) or long ($c_S$) path at PBS1 and combined again in the same
path at PBS2. The target photon is split along the short ($t_L$) or
long ($t_S$) path at the first beam splitter BS1 and combined again
in the same path at BS2. A HWP oriented at $45^{\circ}$ rotates the
polarization state of the photon taking the long path by
$90^{\circ}$. The path-length difference $\Delta L$ of $c_L$ and
$c_S$ is the same with that of $t_L$ and $t_S$ and satisfies the
condition
\begin{equation}
l_{\text{SPDC}}\leqslant\Delta L\leqslant l_{\text{pump}},
\end{equation}
where $l_{\text{SPDC}}$ is the coherence length of the
down-converted photon and $l_{\text{pump}}$ is the spectral width of
the pump laser. Conditioned on coincidence of detection with the
time window of the coincidence counter satisfying \mbox{$\Delta
T<\Delta L/c$}, we can write the evolution of an arbitrary input
state as
\begin{align}\label{CNOT in Sanaka}
b_1\ket{H}_{c_{\text{in}}}\ket{H}_{t_{\text{in}}}+&b_2\ket{H}_{c_{\text{in}}}\ket{V}_{t_{\text{in}}}+b_3\ket{V}_{c_{\text{in}}}\ket{H}_{t_{\text{in}}}\nonumber\\
+b_4\ket{V}_{c_{\text{in}}}\ket{V}_{t_{\text{in}}}&\longrightarrow\nonumber\\
b_1\ket{H^S}_{c_{\text{out}}}&\ket{H^S}_{t_{\text{out}}}+b_2\ket{H^S}_{c_{\text{out}}}\ket{V^S}_{t_{\text{out}}}\nonumber\\
+b_3\ket{V^L}_{c_{\text{out}}}&\ket{V^L}_{t_{\text{out}}}+b_4\ket{V^L}_{c_{\text{out}}}\ket{H^L}_{t_{\text{out}}},
\end{align}
where $b_i (i=1,2,3,4)$ is an arbitrary complex number satisfying
normalization condition, and the superscript $S$ ($L$) denotes the
photon passing the short (long) path. Here the coincidence counting
has post-selected out unwanted state components in which the control
and target photons followed paths of different lengths. Because of
the time-energy entanglement, paths of the same length are
indistinguishable and so add coherently. The success probability is
$1/4$.

\begin{figure}[tb]
\centering
\includegraphics[width=8.5cm]{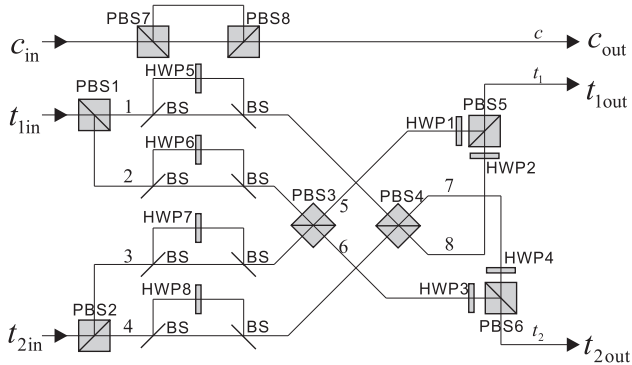}
\caption{Optical realization of a post-selected linear optical
Fredkin gate assisted by time entanglement. Lowercase letters and
numbers label the beams. Photons in beams $c_{\text{in}}$,
$t_{1\text{in}}$ and $t_{2\text{in}}$ are the control and two target
qubits. Polarizing beam splitters (PBS) transmit horizontally
polarized photons and reflect vertically polarized photons. BS is
balanced beam splitter. Half-wave plates HWP1 and HWP3 are oriented
at $67.5^{\circ}$. HWP2 and HWP4 are set to $22.5^{\circ}$. HWP5,
HWP6, HWP7 and HWP8 are set to $45^{\circ}$. The gate succeeds in
the case of threefold coincidence detection in the output modes $c$,
$t_1$ and $t_2$.} \label{fig_timeCNOT}
\end{figure}

Fig.~\ref{fig_timeCNOT} shows an optical realization of a
post-selected Fredkin gate by replacing the four CNOT gates in
Fig.~\ref{fig_fourCNOT} with the CNOT gates we have just introduced.
Based on the analysis above, in the case of the input state given by
Eq.~(\ref{input}), the successful output state can be found to be
\begin{align}\label{output_Time}
&a_1\ket{H^S}_c\ket{H^S}_{t_1}\ket{H^S}_{t_2}+a_2\ket{H^S}_c\ket{H^S}_{t_1}\ket{V^S}_{t_2}\nonumber\\+&a_3\ket{H^S}_c\ket{V^S}_{t_1}\ket{H^S}_{t_2}+a_4\ket{H^S}_c\ket{V^S}_{t_1}\ket{V^S}_{t_2}\nonumber\\
+&a_5\ket{V^L}_c\ket{H^L}_{t_1}\ket{H^L}_{t_2}+a_6\ket{V^L}_c\ket{V^L}_{t_1}\ket{H^L}_{t_2}\nonumber\\+&a_7\ket{V^L}_c\ket{H^L}_{t_1}\ket{V^L}_{t_2}+a_8\ket{V^L}_c\ket{V^L}_{t_1}\ket{V^L}_{t_2}.
\end{align}

From Eqs.~(\ref{CNOT in Sanaka}) and (\ref{output_Time}), we can see
that to make the output state entangled the three input photons need
be time-entangled in the two time bins, ``$S$'' and ``$L$''. This
scheme needs no ancillary photons and the probability of success is
$1/64$. Three qubit time entangled states of the type required, i.e.
in which a triple coincidence is in a superposition of many times,
have been described in \cite{Keller_three, hnilo_three, Anno_multi},
however, an experimental demonstration of such states has not yet
been made.

\section{Conclusions}\label{Sec_conclusion}

We have discussed the implementation of Fredkin gate with linear
optics and single photons. We have presented a heralded method using
four heralded CNOT gates. Our method needs less ancillary photons
than that in \cite{Smolin1996}, but is less efficient than
Fiur\'{a}\v{s}ek's scheme \cite{fiurasek_fredkin}. We have also
simplified the heralded scheme to a post-selected one by replacing
two CNOT gates with two HWPs. This scheme needs only two ancillary
photons, and therefore is feasible with existing technology.
However, the low success probability would make the experiment more
difficult. The other post-selected Fredkin gate we have proposed is
assisted by time entanglement. Although this scheme needs no
ancillary photons, the three-photon time entangled source required
is not available at present.

It should be noted that since the post-selected schemes work in the
coincidence basis, such schemes could not be scalable unless
photon-number Quantum nondemolition (QND) detectors were added to
each output beam, nevertheless they open the door for the first time
to experimental tests of an optical Fredkin gate and would make its
application possible. We hope our proposals will stimulate such
investigations of Fredkin gate.

\begin{acknowledgments}
YXG thanks Yun-Feng Huang for useful discussions. YXG and GCG were
funded by National Fundamental Research Program (Grant No.
2006CB921907), National Natural Science Foundation of China (Grant
No. 60121503 and No. 60621064), Innovation Funds from Chinese
Academy of Sciences, International Cooperate Program from CAS and
Ministry of Science \& Technology of China. TCR was supported by the
DTO-funded U.S.Army Research Office Contract No. W911NF-05-0397 and
the Australian Research Council.
\end{acknowledgments}

\bibliography{Fredkinref}

\end{document}